\documentclass[aps,twocolumn,amsfonts,amsmath,amssymb,prb]{revtex4}
\pdfoutput=1
\usepackage{graphicx}
\usepackage{graphics}
\newcommand*{\revrem}[1]{}
\usepackage{amsfonts}
\usepackage{amssymb}
\usepackage{bm}

\begin{document}

\title
  {Confinement Effects on an Electron Transfer Reaction in Nanoporous Carbon Electrodes}

\author{Zhujie Li$^{1,2,3}$, Guillaume Jeanmairet$^{2,3}$, Trinidad M\'{e}ndez-Morales$^{1,2,3}$, Mario Burbano$^{1,3}$, Matthieu Haefele$^1$, Mathieu Salanne$^{1,2,3}$}
\affiliation
{$^1$Maison de la Simulation, CEA, CNRS, Univ. Paris-Sud, UVSQ, Universit\'{e} Paris-Saclay, F-91191 Gif-sur-Yvette, France}
\affiliation
{$^2$Sorbonne Universit\'{e}s, UPMC Univ Paris 06, CNRS, Laboratoire PHENIX, F-75005 Paris, France}
\affiliation
{$^3$R\'eseau sur le Stockage \'Electrochimique de l'\'Energie (RS2E), FR CNRS 3459, France}

\email{mathieu.salanne@upmc.fr}

%\usepackage{color}
%\usepackage{soul}
%\newcommand*{\revision}[2]{{\color{red}{#1}} \textst{#2}}
%\newcommand*{\revadd}[1]{{\color{red}{#1}}}
%\newcommand*{\revrem}[1]{\textst{#1}}

%%%%%%%%%%%%%%%%%%%%%%%%%%%%%%%%%%%%%%%%%%%%%%%%%%%%%%%%%%%%%%%%%%%%%
%% The document title should be given as usual. Some journals require
%% a running title from the author: this should be supplied as an
%% optional argument to \title.
%%%%%%%%%%%%%%%%%%%%%%%%%%%%%%%%%%%%%%%%%%%%%%%%%%%%%%%%%%%%%%%%%%%%%
 
%%%%%%%%%%%%%%%%%%%%%%%%%%%%%%%%%%%%%%%%%%%%%%%%%%%%%%%%%%%%%%%%%%%%%
%% Some journals require a list of abbreviations or keywords to be
%% supplied. These should be set up here, and will be printed after
%% the title and author information, if needed.
%%%%%%%%%%%%%%%%%%%%%%%%%%%%%%%%%%%%%%%%%%%%%%%%%%%%%%%%%%%%%%%%%%%%%
%\abbreviations{} %eg.{IR,NMR,UV}
%\keywords{nanoporous material, ionic liquids, nanoconfinement
%effect, molecular dynamics simulation, redox species, electron transfer
%reaction, Marcus theory}

%%%%%%%%%%%%%%%%%%%%%%%%%%%%%%%%%%%%%%%%%%%%%%%%%%%%%%%%%%%%%%%%%%%%%
%% The manuscript does not need to include \maketitle, which is
%% executed automatically.
%%%%%%%%%%%%%%%%%%%%%%%%%%%%%%%%%%%%%%%%%%%%%%%%%%%%%%%%%%%%%%%%%%%%%

%\makeatother

%\begin{document}
\begin{abstract}
Nanoconfinement generally leads to drastic effect on the physical and chemical
properties of ionic liquids. Here we investigate how the electrochemical reactivity 
in such media may be impacted inside nanoporous carbon electrodes. To this end, we study a simple electron transfer reaction using molecular dynamics simulations. The electrodes are held at constant electric potential by allowing the atomic charges on the carbon atoms to fluctuate. We show that the $\mathrm{Fe^{3+}}/\mathrm{Fe^{2+}}$ couple dissolved in an ionic
liquid  exhibits a deviation with respect to Marcus theory. This behavior is rationalized by the stabilization of a solvation state
of the Fe$^{3+}$ cation in the disordered nanoporous electrode that
is not observed in the bulk. The simulation results are fitted with a recently proposed two solvation state model, which allows us to estimate the effect of such a deviation on the kinetics of electron transfer inside nanoporous electrodes.

\end{abstract}

\maketitle

%%\begin{tocentry}
%\noindent TOC graphic:\\
%%\begin{figure}[ht!]
%\begin{center}
% \includegraphics[width=5cm]{figures/TOC}
%\end{center}
%%\end{tocentry}

%\section*{Introduction}

Nanoconfinement effects
strongly impact on many liquid properties, such as transport, diffusion
coefficients, phase transitions, and solvation structures~\cite{0953-8984-18-6-R01,jiang2013b,Pean2015,agrawal_observation_2016}.
They are particularly important for supercapacitors, which have emerged as a complimentary energy storage solution to  % have emerged
batteries~\cite{salanne2016a}. This is due to the fact that supercapacitors display faster charging times, and consequently higher power deliveries than batteries, while attaining longer life cycles. However, the energy density of batteries is higher.
In supercapacitors energy storage is realized through the adsorption of ions at the surface
of two oppositely polarized electrodes, forming a so-called electrical
double layer \cite{kornyshev2007a}. Nanoconfinement
has been observed to have a significant influence on the performance of supercapacitors, as has been 
demonstrated by experiments\cite{Chmiola2006a,Raymundo-Pinero2006} and simulations\cite{kondrat2012a,merlet_molecular_2012,he2015a,vatamanu2015b}, where it was shown that the use of materials
with sub nanometric pores as electrodes greatly increase the capacitance
of these devices.

Within the realm of electrochemical applications, room temperature ionic liquids
(RTILs) have attracted a considerable attention~\cite{Silvester2006,Armand2009}. In principle, the properties
of RTILs are tunable due to the wide variety of cations and anions
from which they can be prepared. Because RTILs are solely made of charged species, it would have been expected
that using them to replace standard electrolytes would increase
the performance of supercapacitors. Yet, the interfacial capacitance
of RTILs and acetonitrile-based electrolyte supercapacitors remain somewhat
similar. This rather disappointing observation has been attributed
to the more important correlations between ions in RTILs, which is due to the absence of electrostatic  screening
by the solvent~\cite{burt2016a}.

An alternative could be to take advantage of the tunability of RTILs
to develop new storage concepts. Recently, Mourad \textit{et al.} reported % italics for et al.
a large enhancement of the energy stored due to simultaneous capacitive
and Faradic processes when biredox RTILs are used as electrolytes
in a supercapacitor~\cite{Mourad2016}. Among the various questions
raised by this study, the most important ones are: How is the electron
transfer rate affected by confinement? Does the charging of such
a supercapacitor remain dominated by the ionic diffusion? 

From the theoretical point of view, electron transfer reactions in
solution are usually studied in the framework of Marcus theory, which
 aims at accounting for the influence of solvent fluctuations
on the rate of electron transfer~\cite{Marcus1956}. Marcus theory
has been widely used to interpret experiments and simulations with
an undeniable success. However, some systems exhibiting a deviation
from Marcus linear behavior have been reported \cite{small_theory_2003,Blumberger2008}.
Several extensions to the theory were proposed to account
for them~\cite{matyushov_modeling_2000,Vuilleumier2012}. A
reason why Marcus theory might fail arises from the fact that
one of its key assumptions is that the fluctuations of the solvent around
the reactant and the product are similar. Yet, if the solvation states
of the two species are structurally  different, this hypothesis % really different --> structuraly and dyanmically different
can be wrong as evidenced by the use of density functional theory-based Molecular Dynamics
simulations of aqueous copper and silver ions~\cite{Blumberger2008,Vuilleumier2012}. In experiments as well, introducing asymmetry in Marcus theory~\cite{laborda2013a} was necessary to fit the voltammetric data of various RTILs~\cite{tanner2015a}. Finally, in the case of interfacial systems several studies have underlined the importance of field penetration into the metal and of solvent spatial correlations.~\cite{dzhavakhidze1987a,phelps1990a}

Confinement might have a drastic influence on the solvation shell
of the adsorbed species~\cite{Merlet2013,prehal2017a}. Thus, it is worth investigating this effect
on the electron transfer rate~\cite{bai2014a}. 
 We report here a molecular dynamics study of the
Fe$^{3+}$/Fe$^{2+}$ electron transfer reaction which has already been proposed in previous redox supercapacitor concepts~\cite{akinwolemiwa2015a,su2009b}. The experimental studies involve complex species, so that we focus on a simplified system in order to gain a first insight on electron transfer reactions in
carbon nanopores. In particular, our study corresponds to infinite dilution since it is known that inorganic salts have low solubilities in imidazolium-based RTILs~\cite{pereiro2012a}.
 The studied ionic liquid is the 1-ethyl-3-methylimidazolium
tetrafluoroborate (EMIM-BF$_{4}$), which is put in contact with
model carbide-derived carbon (CDC) nanoporous electrodes. We show that the free energy profiles for the electron transfer reaction strongly deviate from Marcus theory due to the presence of two solvation states for the Fe$^{3+}$ species.

%\section{Results and Discussion}

\begin{figure*}[htb]
\centering \includegraphics[width=0.9\textwidth]{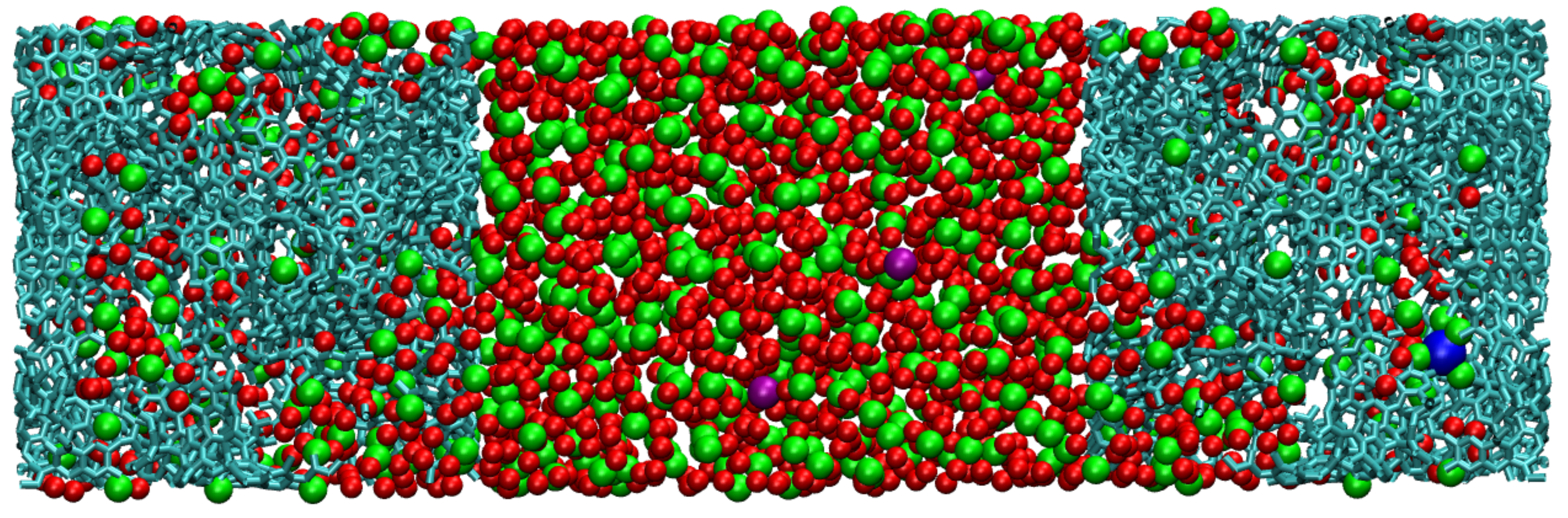}
\caption{The simulation cell is constituted of two porous electrodes held at
constant potential (cyan C atoms). The electrolyte is composed of a FeCl$_3$ or FeCl$_2$ unit dissolved in EMIM-BF$_{4}$, modeled using a coarse-grained force-field\cite{Merlet2012a} (red: the three interaction sites of EMIM$^{+}$, green: the single site of BF$_{4}^{-}$,
blue: Fe (III) cation and violet: Cl$^-$ anions). }
\label{fgr:CDC_stucture} 
\end{figure*}

\begin{figure*}[htb]
\centering \includegraphics[width=0.9\textwidth]{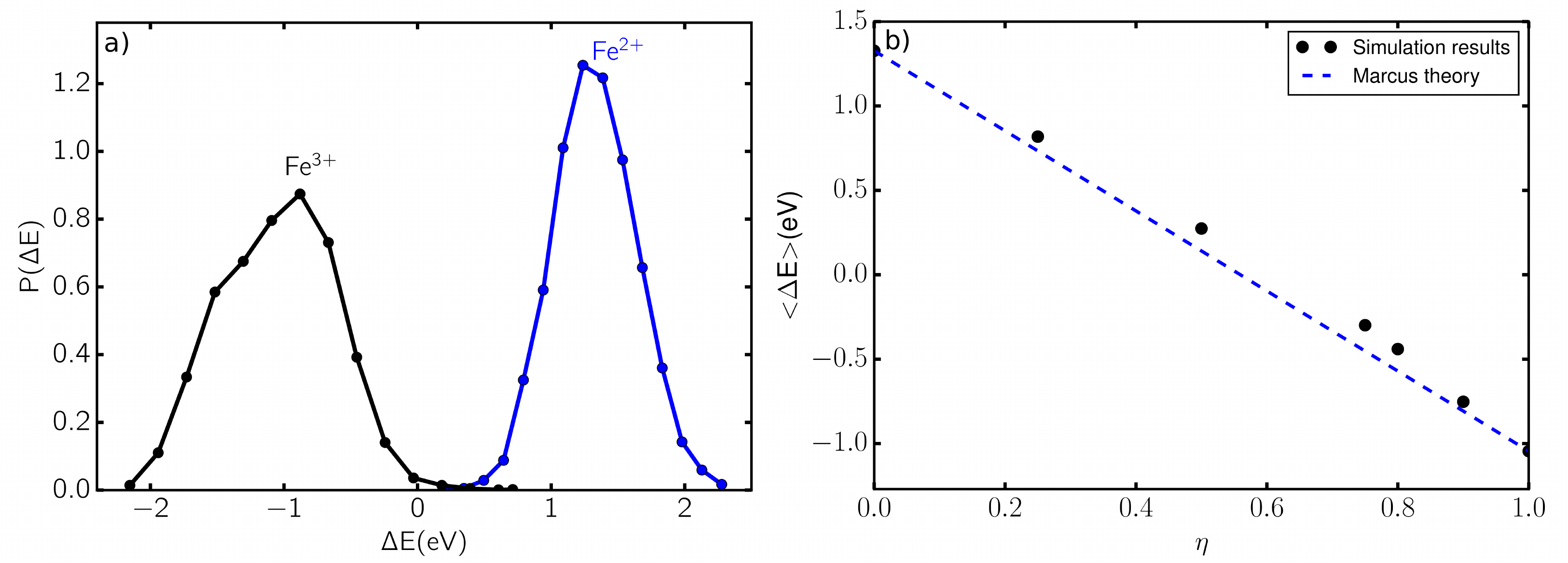}
\caption{a) Probability distribution of the vertical energy gap in reduced
and oxidized states, b) Average vertical energy gap for Fe$^{3+}$/Fe$^{2+}$
in RTILs with various coupling parameters $\eta$. }
\label{fgr:non-Gaussian} 
\end{figure*}

We build on previous work by using a simulation cell similar to the one we used
to investigate the origin of increased capacitance in nanoporous carbon-based supercapacitors\cite{merlet_molecular_2012}. We employ a coarse-grained
model of the EMIM-BF$_{4}$ with, respectively, 3 and 1 interaction sites for the
cation and the anion \cite{roy_improved_2010}, to which we add one
iron ion  and % MB how about the interactions for the Fe ions?
the appropriate number of chloride counter ions. 
We keep the electric potential inside the electrodes  constant and equal to 0$\ $V by using
the procedure developed by Sprik \textit{et al.}\cite{siepmann_influence_1995}\textit{,
} which allows the charge on the electrode to adjust in
response to the local electric potential due to the electrolyte ions. Each electrode
is represented by a model of CDC containing
3821 carbon atoms\cite{Palmer2010}. To address the effect of the
local environment experienced by the redox species we ran several simulations
in which the initial position of the iron ion is set in different
pores of the disordered carbon material. In order to study the following redox half-reaction:
\[
\mathrm{Fe^{2+}\to Fe^{3+}+e^{-}}
\]

\noindent we performed simulations of the reduced species $\mathrm{Fe^{2+}}$ and
of the oxidized one Fe$^{3+}$. It has been shown by Warshel\cite{Warshel1982,Hwang1987}
that a relevant reaction coordinate to study electron transfer reaction
is the vertical energy gap $\Delta E$,  which is defined as

\begin{equation}
\Delta E(\{{\bf{R}}^N\})=E_{1}(\{{\bf{R}}^N\})-E_{0}(\{{\bf{R}}^N\}),\label{eq:delta_E}
\end{equation}

where $E_{1}$ and $E_{0}$ are the instantaneous potential energies of a system with either the reduced or the  
oxidized state of the redox-active species for a given microscopic configuration $\{{\bf{R}}^N\}$.
We recall here that in the linear response assumption in Marcus theory \cite{Marcus1956,Marcus1965}, the distribution of
the order parameter is Gaussian\cite{Georgievskii1999}
and, in particular, the shape of the distribution of $\Delta E$ obtained
by performing simulations with the Hamiltonian of the reactant or the
product should be identical. As shown in \ref{fgr:non-Gaussian}a,
this is clearly not the case, the distribution  for the Fe$^{2+}$ being almost Gaussian, while the
one for the Fe$^{3+}$ is not. This is a first indicator % MB evidence --> indicator
that this system does not follow Marcus theory picture.

In order to increase the statistical accuracy, we use the free-energy perturbation method. This variant of umbrella sampling introduces a coupling parameter ($\eta$) between the reduced and oxidized states~\cite{King1990}. The simulations are then performed with an intermediate Hamiltonian
 associated to the potential energy surface
 $E_{\eta}$ defined as the linear superposition:  % MB moved the end of the sentence here to make it more readable. 

\begin{equation}
E_{\eta}=(1-\eta)E_{0}+\eta E_{1}\label{eq:PES}
\end{equation}

\noindent A direct consequence of the linear assumption of Marcus theory is that $E_{\eta}$ should vary linearly with $\eta$\cite{King1990}. To test this, we carried out simulations
with different values of the coupling parameter $\eta$ = 0, 0.25, 0.5,
0.75, 0.8, 0.9 and 1.0.  It is clear for \ref{fgr:non-Gaussian}b that the
Marcus picture is violated, since the corresponding vertical energy gaps do not follow the expected linear variation.

%\subsection{2. The coordination structure}
\begin{figure*}[htb]
\centering \includegraphics[width=0.8\textwidth]{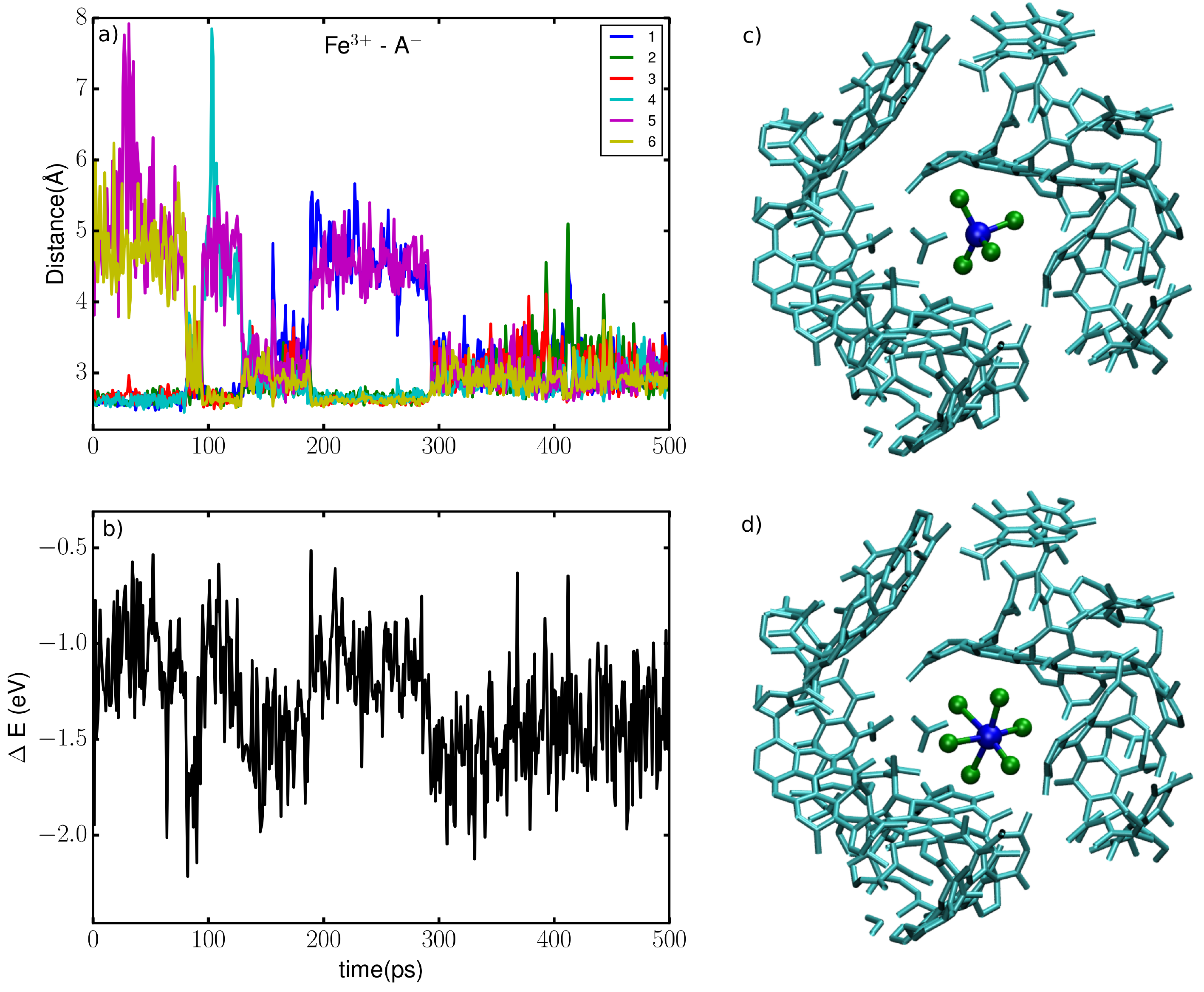}
\caption{a) Distance between Fe$^{3+}$ and the six closest anions, b)
Vertical energy gaps of the oxidized state Fe$^{3+}$, c) and d) Snapshots
of the two different solvation structures.}
\label{fgr:cn} 
\end{figure*}

As pointed out in the introduction, discrepancy with Marcus theory
often originates from strong changes in the solvation shell between
the reduced and oxidized states. To investigate if that is indeed
the case here, we examine the coordination number
of the iron cation along the simulations. The radius of the first solvation shell is taken equal to the first minimum of the Fe-BF$_{4}^{-}$
radial distribution function in the absence of electrodes (3.8~\AA). We observe
that the coordination number (CN) of the Fe$^{3+}$ cation fluctuates between values of 4 and 6
during the simulation. This is clearly an effect of the confinement
since the cation remains close to the carbon electrode, and the corresponding CN
 is 4 in the pure RTIL. In contrast,
the Fe$^{2+}$ cation has a CN that remains equal to 4 during the
whole simulations, as shown in Supplementary Figure S1.

The distances between
Fe$^{3+}$ and the six nearest anions along a representative trajectory are displayed in \ref{fgr:cn}a;
the other simulations display similar variations. The alternation between two coordination environments is confirmed by the fact that during some parts of
the simulations we detected that 4 anions are closely coordinated
to the Fe$^{3+}$ with a Fe-BF$_{4}$ distance of 2.8 \AA,
the two remaining anions being located further away from the iron with
a distance greater than 4 \AA, while in other parts 
the 6 anions are coordinated with Fe$^{3+}$ with metal-ligand distances
fluctuating between 2.6 and 3.8 \AA. \ref{fgr:cn}c
and \ref{fgr:cn}d show representative snapshots
of the two coordination states which appear to be tetrahedral and
octahedral, respectively. The corresponding fluctuations of the vertical energy gap are shown on \ref{fgr:cn}b. Sudden changes occur concommitently with 
the jumps in CN in \ref{fgr:cn}a, which shows that the confinement effect on the solvation of Fe$^{3+}$ is at the origin of the departure from Marcus
theory discussed above.

Among all models available in the literature
to extend Marcus theory to non-linear behavior~\cite{blumberger2015a} we use the one proposed
by Vuilleumier \textit{et al.} \cite{Vuilleumier2012}. In particular, we employed % picked --> employed
their two-Gaussian solvation (TGS) state model, which allows the reactant
and product to experience two different solvation states (S$_{0}$
and S$_{1}$). This leads to four (two pairs) diabatic free energy
parabolas instead of two in the case of Marcus theory. For each of
the states S$_{0}$ or S$_{1}$, the Landau free energy is assumed
to be quadratic, which corresponds to a Gaussian probability distribution
$P_{\eta}$. The authors also derived the necessary equations to compute
the average energy gap $\left\langle \Delta E\right\rangle _{\eta}$
and the Landau free energies.

 Marcus theory involves only 2 parameters: $\Delta A$ measures the relative position of the
two parabolas while $\lambda$ is the reorganization energy of the solvent, which fixes
the curvature of the parabola and also the width of the Gaussian
distribution of the reaction coordinate. In  the TGS model,
4 parameters are required, one reorganization energy for each state
$\lambda_{S_{0}}$ and $\lambda_{S_{1}}$ and the relative positions
of the two parabola for each solvation state $\Delta A_{S_{0}}$ and
$\Delta A_{S_{1}}$. Moreover, a fifth parameter is necessary to set
the relative position of the two pairs of parabolas, for instance the
difference in free energy between the two solvation states of the
reduced species $\Delta_{S}A_{0}=A^{\rm Fe^{2+}}_{S_{1}}-A^{\rm Fe^{2+}}_{S_{0}}$.

The effect of the confinement was not investigated by the authors
of the TGS model, since they only carried out simulations in bulk.
However, it has been shown from simulation studies of the $\mathrm{Eu^{3+}/Eu^{2+}}$
redox couple in potassium chloride molten salt at an fcc metallic
electrode that the reorganization energy strongly depends on the distance
between the ion and the electrode~\cite{Pounds2015}. This reorganization
energy is expected to decrease when the redox species get closer to
the electrode. We mention that this behavior had previously been predicted
theoretically by Marcus for perfect conductors,\cite{Marcus1965,marcus_reorganization_1990} but this effect should decrease if the role of field penetration into the metal electrode was accounted.\cite{dzhavakhidze1987a,phelps1990a} 

To quantify this effect in our system, we
performed additional simulations
with a planar graphite electrodes. They confirm the dependency on the
distance to the electrode of the reorganization energy, as shown in Supplementary Figures
S2 to S4. In nanoporous electrodes it is not possible to define a simple distance to the electrode, the fit of the TGS model should therefore yield an averaged contribution
of all the possible distances. However
we observed that the iron does not go into contact with the carbon during the whole simulation data, regardless of the iron solvation
state, the nature of the pore in which it is solvated or its coordination
number. For this reason and for the sake of simplicity we made the
additional assumption that the reorganization energies should be equal
for both states, turning the original 5-parameter model TGS into
a 4 parameter-one. The different parameters obtained by fitting the simulation
 data are given in \ref{tbl:1}.\\

\begin{table}
\caption{TGS model parameters for Fe$^{3+}$/Fe$^{2+}$ obtained by fitting
simultaneously on $P_{\eta}(\Delta E)$ and $\left\langle \Delta E\right\rangle _{\eta}$.}
\label{tbl:1} %
\begin{tabular}{llllll}
\hline 
TGS parameters & $\lambda_{S_{0}}$  & $\Delta A_{S_{0}}$  & $\lambda_{S_{1}}$  & $\Delta A_{S_{1}}$  & $\Delta_{S}A_{0}$ \tabularnewline
\hline 
(eV)  & 1.11  & 0.24  & 1.11  & -0.27  & 0.52\tabularnewline
\hline 
\end{tabular}
\end{table}

\begin{figure*}[htb]
\centering \includegraphics[width=0.9\textwidth]{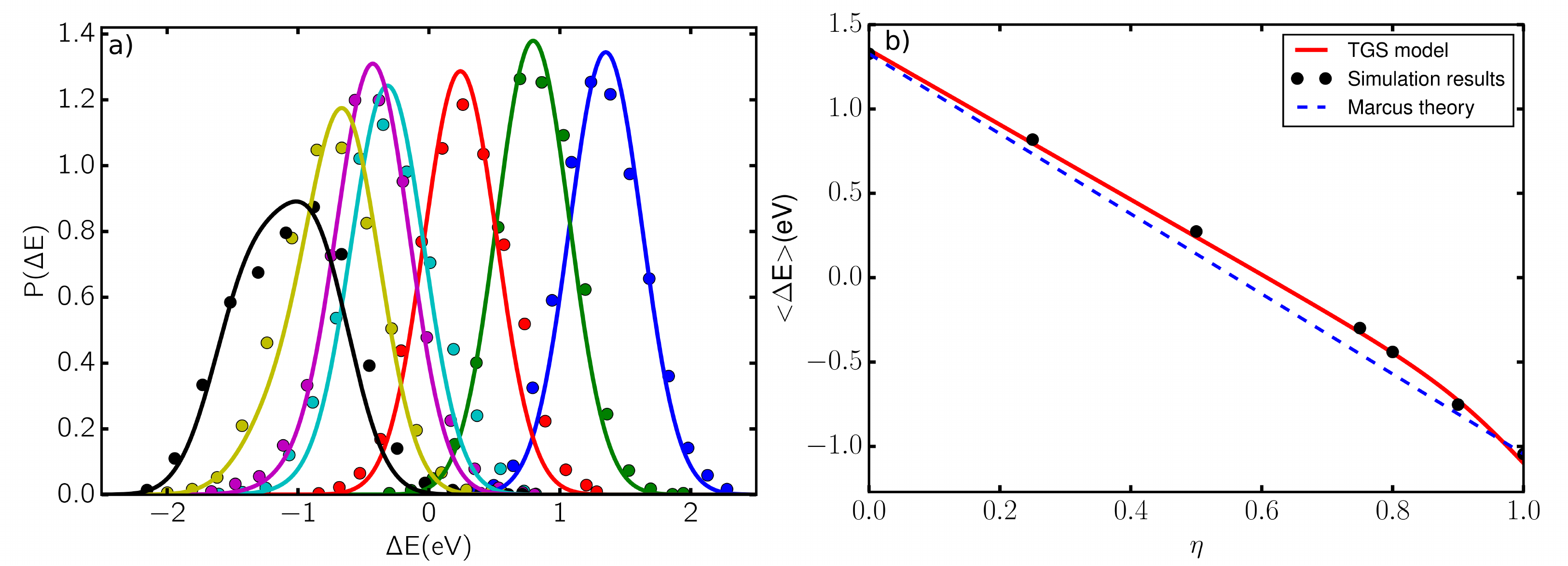}
\caption{a) Probability distribution of the vertical energy gap,
as a function of the coupling parameter $\eta$. Points are the
simulation data, and lines are the TGS model with the parameters from \ref{tbl:1}. b) Comparison between the simulated average vertical energy gap for Fe$^{3+}$/Fe$^{2+}$
in RTILs with various coupling parameters $\eta$ (points) and 
the TGS model (red line).}
\label{fgr:TGS_probability_lines} 
\end{figure*}

\ref{fgr:TGS_probability_lines} compares the
results obtained by simulations to the TGS model with the parameters
given in \ref{tbl:1}. The left hand panel shows the probability
distribution of the vertical energy gap, for the simulations realized
with different values of the $\eta$ parameter. The fitted TGS
model agrees well with the whole set of distribution. The right hand panel reproduces
the data of \ref{fgr:non-Gaussian} i.e. the vertical energy
gap as a function
of $\eta$. While those data could not be well reproduced by assuming
 linear response, the use of the TGS model allows
an almost perfect fit to the data.

%\subsection{4. The mechanism digram of the ET reaction}
\begin{figure}[htb]
\centering \includegraphics[width=\columnwidth]{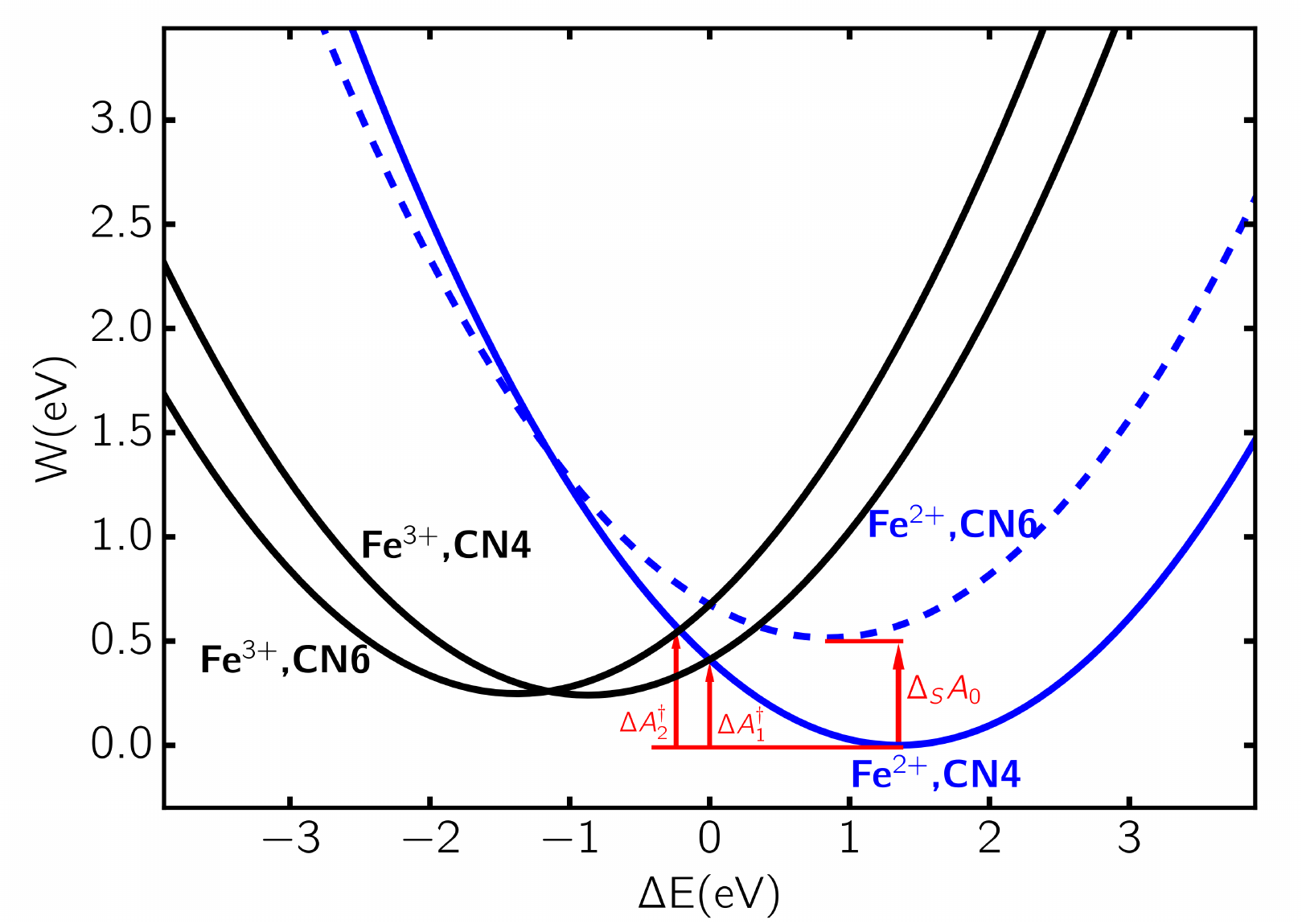}
\caption{Diabatic free energy curves of the reduced and oxidized species obtained
by TGS model with the parameters of \ref{tbl:1}. ``CN4''
and ``CN6'' labels refer to the tetra and
hexacoordinated solvation states and correspond to S$_{0}$
and S$_{1}$ respectively.}
\label{fgr:TGS_parabola} 
\end{figure}

The diabatic free energy curves for the reduced and oxidized species
in Fe$^{3+}$/Fe$^{2+}$ predicted by the TGS model with the parameters
of \ref{tbl:1} are plotted in \ref{fgr:TGS_parabola}.
The two parabolas for the reduced or oxidized
state ($\eta$ = 0 or 1) correspond to a given solvation state.
The free energy difference ($\Delta_{S}A_{0}$)
between the hexacoordinated  and the tetracoordinated forms of Fe$^{2+}$
is large (0.52 eV) with respect
to the thermal energy $k_B T$ (34.5~meV at $T~=~400$~K). The hexacoordinated state is therefore very unlikely, which explains why it is never observed in our equilibrium simulations. On the other hand,
the free energies of the two coordination states are
very close from each other for Fe$^{3+}$. 

\begin{figure}[htb]
\centering \includegraphics[width=\columnwidth]{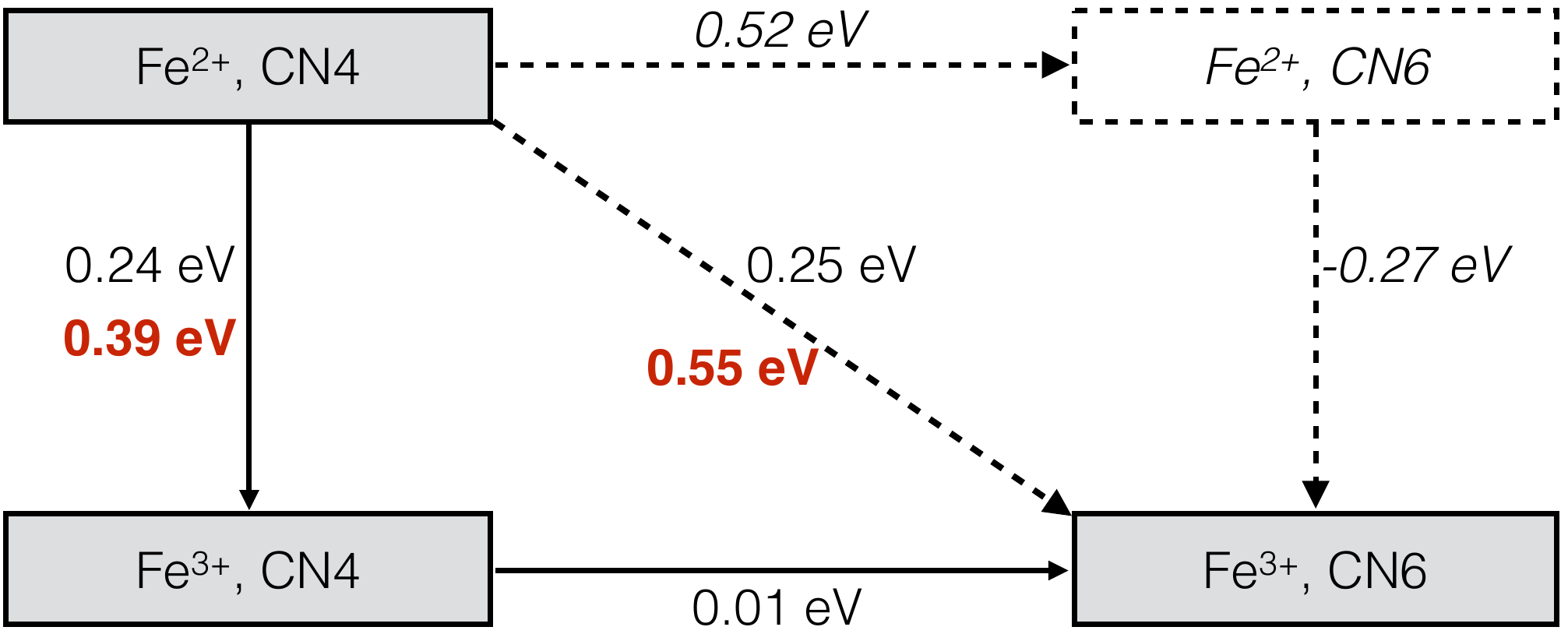}
\caption{Mechanistic diagram of the electron transfer reaction under confinement. Black numbers: free energy differences; red numbers: activation energies (labelled $\Delta A^\dagger_1$ and $\Delta A^\dagger_2$ on \ref{fgr:TGS_parabola}).}
\label{fgr:mechanism} 
\end{figure}

The thermodynamic and kinetic informations extracted from \ref{fgr:TGS_parabola}
are summarized in a mechanistic diagram in \ref{fgr:mechanism}.
We can see that the electron transfer from tetracoordinated
Fe$^{2+}$ to hexacoordinated Fe$^{3+}$ is as likely as the transfer
to the tetracoordinated Fe$^{3+}$ from a thermodynamic point of view,
since the corresponding free energy differences  are very
similar, 0.24 and 0.25 eV respectively. From a kinetic point of view
the picture is completely different since the corresponding activation energy, taken at the intersection between the two parabolas,
 is much higher (0.55~eV vs. 0.39~eV). As a consequence, the effect of the confinement is likely to have a negative impact on the kinetic rate of the electron
transfer reaction. Indeed, it allows for the existence of hexacoordinated Fe$^{3+}$, which is not stable in the bulk liquid in our simulations. The electron
transfer from Fe$^{2+}$ to this chemical species has a very large activation energy barrier, making such events unlikely to occur. This result contrasts with the observation of Remsing\textit{ et al.} who
observed an enhancement of the electron transfer due to the confinement\cite{Remsing2015} in a very different system.
This is easily explained because in their work both the reduced and oxidized species where desolvated with respect to the bulk, while we observe the opposite behavior.

%\section{Conclusion}

In conclusion, the aim of this work was to investigate the free energy properties
of an electron transfer reaction in a nanoporous carbon electrode.
We performed MD simulations  of the Fe$^{3+}$/Fe$^{2+}$ redox couple dissolved
in EMIM-BF$_{4}$ RTIL in contact with CDC
electrodes. The latter were maintained at a constant potential of 0~V 
during the simulations. The vertical energy gap between the redox
species was used as the reaction coordinate, which enabled us to interpret
the microscopic solvent fluctuation properties. The computed probability
distribution of the vertical energy gaps and the equilibrium value
of energy gaps do not follow the linear response approximation
of standard Marcus theory. We demonstrated the strong influence of the nanoconfinement effect
on the solvation shell of the iron cations, which is the main reason for this departure from Marcus theory. In particular,
the Fe$^{3+}$ cation, which is tetracoordinated in the bulk, admits two stable solvation states in the nanoporous material:
tetracoordinated and hexacoordinated. To account for this deviation we used the two-Gaussian solvation model
~\cite{Vuilleumier2012}, from which  the free energy curves for all the redox species in their various solvation states were extracted. This allowed us to qualitatively
analyse the effect of the confinement on the electron transfer reaction.
The fluctuations in the structure of the solvation shell of Fe$^{3+}$ were shown to have a negligible  effect from a thermodynamic point of view. Rather, it was shown that the activation energy of the associated electron transfer process is much higher for the hexacoordinated
form. It is therefore likely that the stabilization of this solvation state will result in a slow down of the electron transfer reaction kinetic.
This work is a first step towards a deeper understanding
of the influence of the confinement on the electron transfer in redox supercapacitor
devices~\cite{akinwolemiwa2015a}, which will be extended in future years to promising systems such as biredox RTILs~\cite{Mourad2016}. To this end, it will be necessary to simulate more complex redox probes and to account systematically for the various complexation states. Another important aspect will be to decouple the contribution of the electron transfer event from the work-term that controls the approach of the ion to the interface in nanoconfinement~\cite{bai2014a}. The techniques develop in this work could also provide useful information for the development of ionic liquids-based thermo-electrochemical cells~\cite{abraham2011a}.\\

\noindent {\bf Methods}\\

We performed molecular dynamics (MD) simulations of the present systems
using coarse-grained force fields. The non-bonded interactions are
represented using Lennard-Jones and Coulomb potentials, and the corresponding
parameters for carbon atoms and EMIM-BF$_4$ interaction sites we used herein are obtained
from our previous work \cite{Merlet2012a,merlet2013b}. The parameters for the chloride and the iron ions are respectively taken from references \cite{Dang1995} and \cite{Ando2001} (the chloride anions were added in order to maintain the electroneutrality of the simulation cell, however their effect on the vertical energy gap was not studied since we focus on infinite dilution conditions). The simulations were conducted
in the NVT ensemble, with the temperature set at 400 K using a Nos\'e-Hoover thermostat (relaxation time: 10~ps). The simulation
cell is orthorhombic, with $x = y = 4.37$~nm, $z = 14.86$~nm,
which reproduces the density of the RTILs compared with
experimental results. Periodic boundary conditions are employed along
$x$ and $y$ directions only through the use of a 2D-Ewald summation~\cite{reed2007a}. For all the simulations, the time step is
2~fs and the time scales of productions are over 500~ps after equilibration. The electrodes are maintained at a constant potential (0~V in our
present study) during the production runs. The vertical energy gap is sampled every 1~ps during the simulation. Following our previous work, a constant has been added to the energies calculated in the simulation to represent the (gas-phase) ionisation potential of Fe$^{2+}$ and the work function of the metal~\cite{Pounds2015}; in practice, the value (15.78~eV) was chosen to bring the calculated redox potential close to the typical experimental values for the Fe$^{3+}$/Fe$^{2+}$ couple in RTILs~\cite{yamato2013a}. 

\newpage

\section*{Supplementary figures}

	\begin{figure}[htb]
		\centering
		\includegraphics[width=\columnwidth]{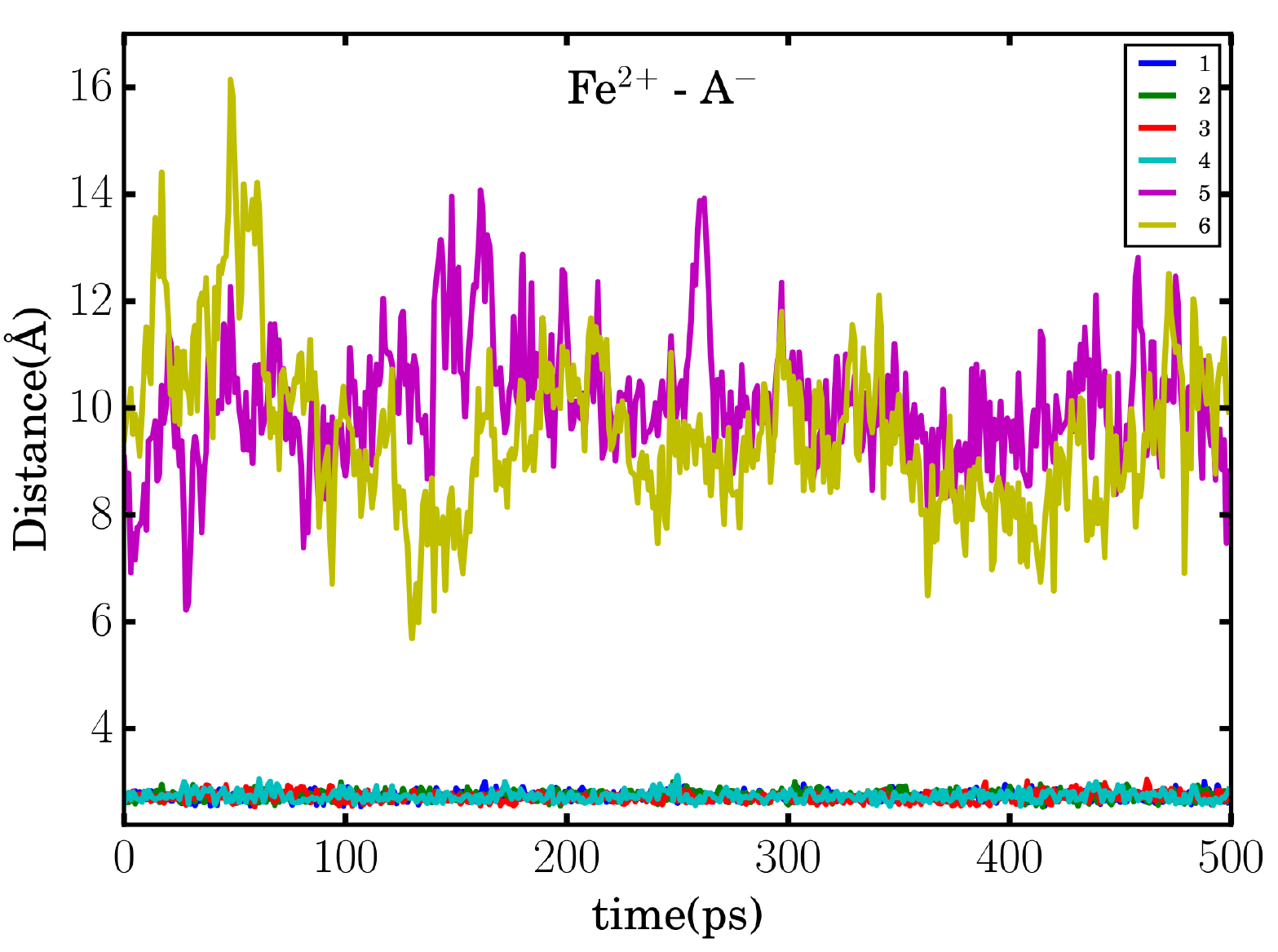}
		\caption{The distance between selected anions (A$^-$) and Fe$^{2+}$ with respect to time, in CDC cell.}
		\label{fgr:cn2}
	\end{figure}
	
	\begin{figure}[htb]
		\centering
		\includegraphics[width=\columnwidth]{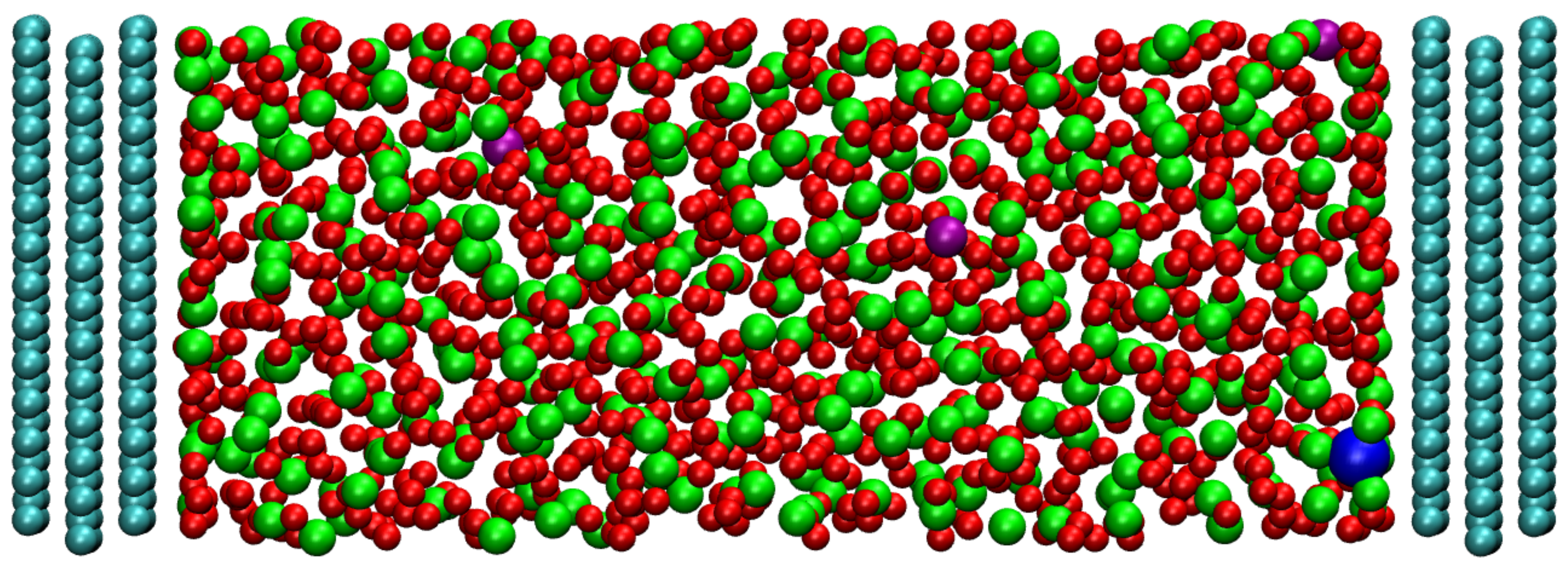}
		\caption{The structure of simulation cell comprising EMIM-BF$_4$ ionic liquids and graphite planar electrodes.}
	\end{figure}

	\begin{figure}[htb]
		\centering
		\includegraphics[width=\columnwidth]{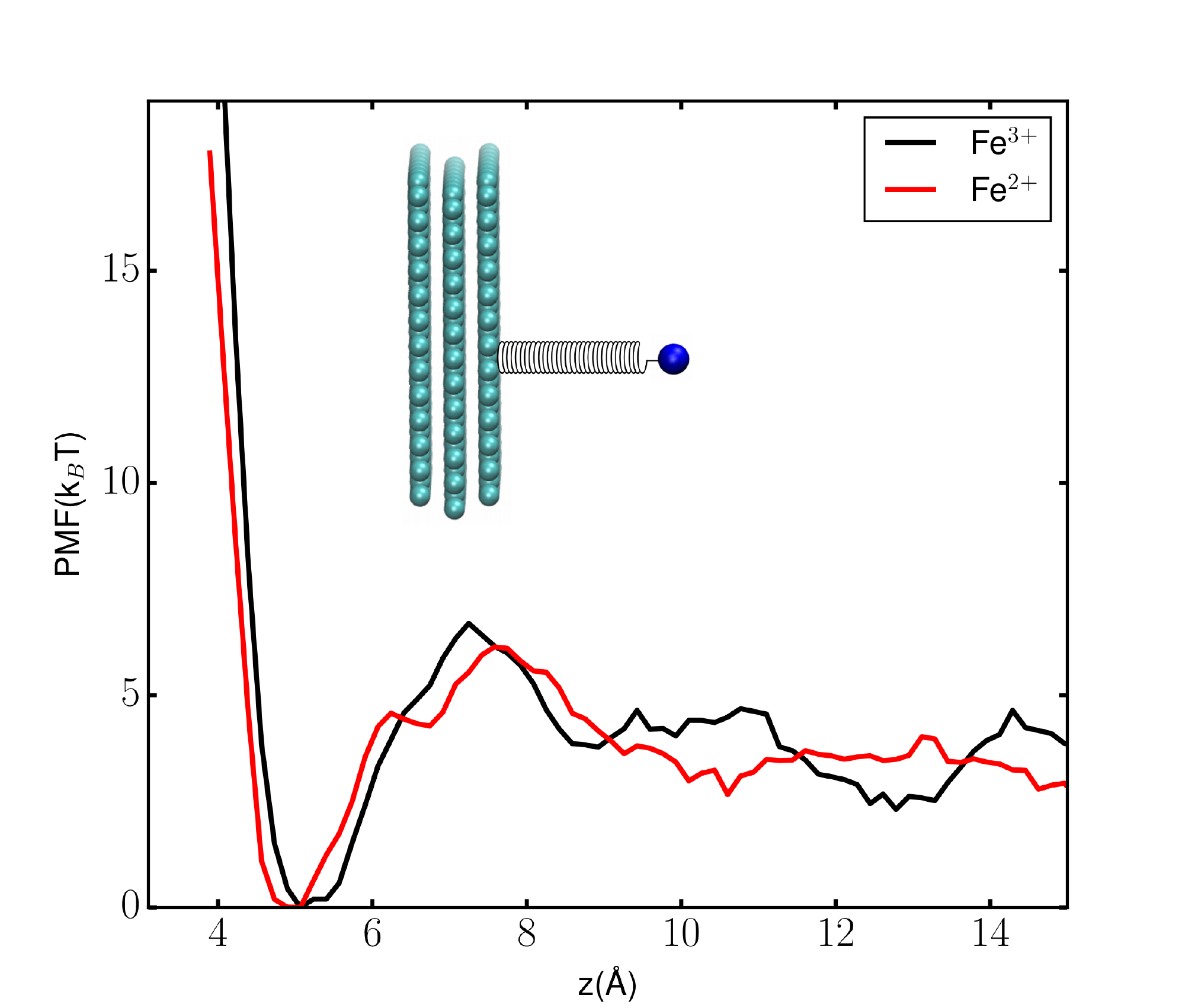}
		\caption{The potentials of mean force for the Fe$^{2+}$ and Fe$^{3+}$  ions approaching the electrode, which are calculated by umbrella sampling method. The z value indicates the distance between iron and electrode as illustrated in the model of umbrella sampling(inset). }
		\label{fgr:PMF}
	\end{figure}

	\begin{figure}[htb]
		\centering
		\includegraphics[width=\columnwidth]{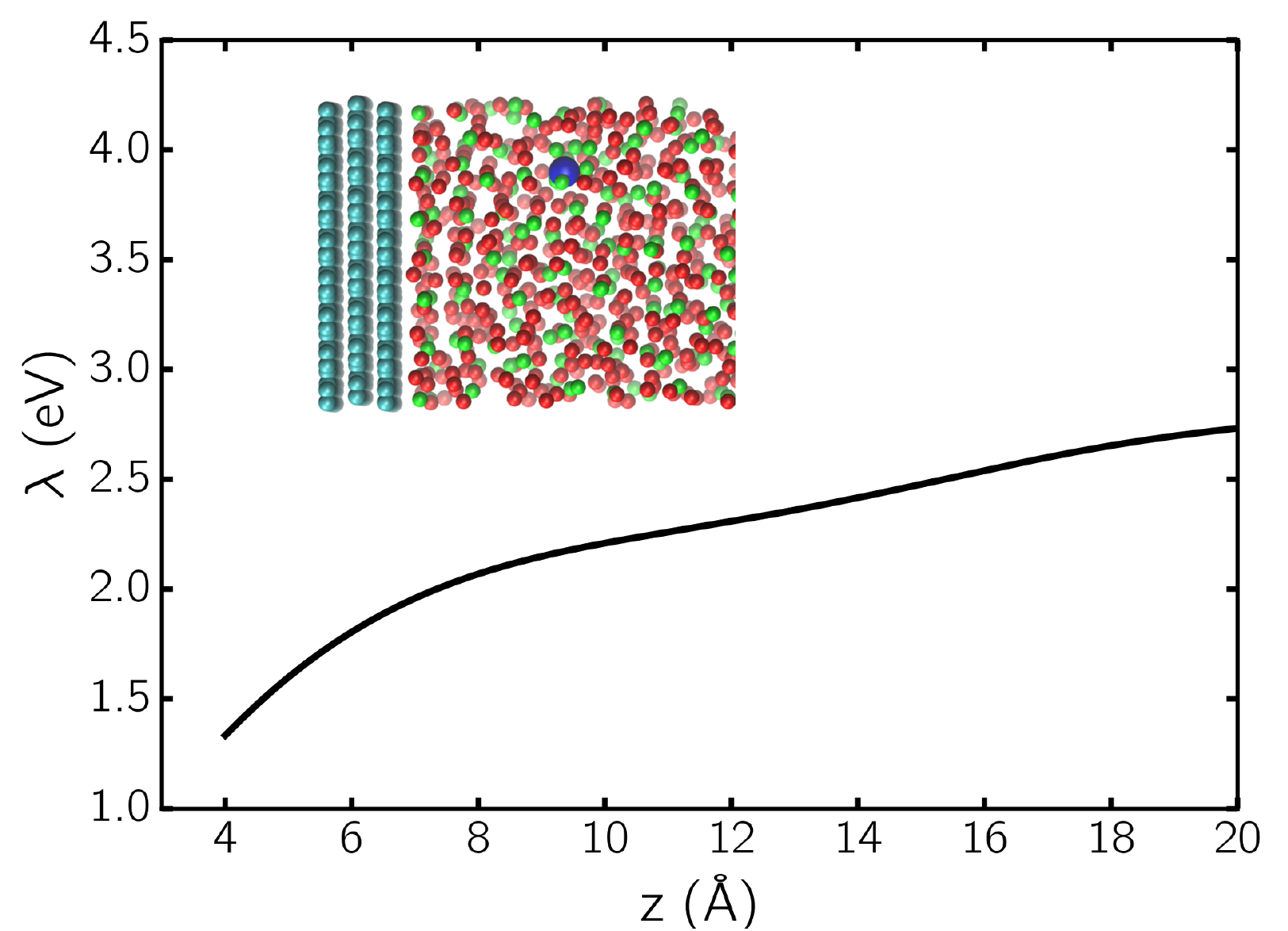}
		\caption{The calculated reorganization energy of the Fe$^{2+}$/Fe$^{3+}$ redox species using Marcus theory, as a function of distance between iron and electrode.}
		\label{fgr:lambda}
	\end{figure}

\section*{acknowledgement}
This work was supported by the French National Research Agency (Labex STORE-EX,  Grant No. ANR-10-LABX-0076). We acknowledge support from EoCoE, a project funded by the European Union Contract No. H2020-EINFRA-2015-1-676629, from the DSM-\'energie programme of CEA and from the Eurotalent programme. We are grateful for the computing resources on OCCIGEN (CINES, French National HPC) and CURIE (TGCC, French National HPC) obtained through the project x2016096728.

%\section*{Associated content}
%{\bf Supporting information available:} Distance between the anions and Fe$^{2+}$ inside the CDC electrode, snapshot of the simulated graphite-based electrochemical cell, potentials of mean force with respect to the distance to the graphite, corresponding reorganization energies (PDF)  

\bibliography{biblio}

\end{document}